\documentstyle[12pt]{article}
\newcommand{\al}{\alpha'}

\newcommand{\de}{\partial}
\newcommand{\db}{\bar{\partial}}

\newcommand{\ca}{\mathcal}
\newcommand{\lr}{\leftrightarrow}

\newcommand{\f}{\frac}
\newcommand{\s}{\sqrt}

\newcommand{\p}{\psi}
\newcommand{\bp}{\tilde{\psi}}

\newcommand{\ap}{\alpha}
\newcommand{\tap}{\tilde{\alpha}}
\newcommand{\bap}{\bar{\alpha}}
\newcommand{\tbap}{\tilde{\bar{\alpha}}}
\newcommand{\be}{\beta}
\newcommand{\tbe}{\tilde{\beta}}
\newcommand{\bbe}{\bar{\beta}}
\newcommand{\tbbe}{\tilde{\bar{\beta}}}
\newcommand{\et}{\eta}
\newcommand{\bet}{\bar{\eta}}
\newcommand{\tet}{\tilde{\eta}}
\newcommand{\tbet}{\tilde{\bar{\eta}}}
\newcommand{\bxi}{\bar{\xi}}
\newcommand{\txi}{\tilde{\xi}}
\newcommand{\tbxi}{\tilde{\bar{\xi}}}

\newcommand{\mb}{\mathbf}
\newcommand{\ddd}{\cdot\cdot\cdot}
\newcommand{\ep}{\epsilon}

\newcommand{\g}{\Gamma}

\newcommand{\fkN}{\frac{k}{N}}
\begin{document}
\thispagestyle{empty}
\begin{flushright}
hep-th/9912157 \\
UT-869\\
December,1999\\
\end{flushright}

\bigskip
\bigskip

\begin{center}
\noindent{\Large String Creation and Monodromy from Fractional D-branes
 on ALE spaces}\\
\bigskip
\bigskip
\noindent{ Tadashi Takayanagi\footnote{
                 E-mail: takayana@hep-th.phys.s.u-tokyo.ac.jp}}\\
\bigskip
{\it Department of Physics, Faculty of Science, University of Tokyo \\
\medskip
Tokyo 113-0033, Japan}
\bigskip
\bigskip
\end{center}
\begin{abstract}

We investigate the anomalous creation of fundamental strings using the boundary
 state formalism of fractional D-branes on ALE spaces in the orbifold limit. 
 The open string Witten index plays a crucial role in this calculation and so the result remains unchanged  even if we blow up the orbifold geometrically, matching the anomaly inflow argument.
  
Further we consider the quiver gauge theories on such fractional D3-branes and see that the string creation mechanism determines 1-loop logarithmic monodromy of these gauge theories. Also we comment on the relation of D(-1)-D3 amplitude to the 1-loop beta function.

\end{abstract}

\newpage

\section{Introduction}
\setcounter{equation}{0}
Since the discovery of D-branes\cite{pol4}, nonperturbative aspects of
superstring theory have been intensively investigated and the tremendous
progress has been made.

Usually, a D-brane is defined as a solitonic object to which open
strings attach. But there is another equivalent description from the
viewpoint of closed strings and it is called the boundary state
\cite{green1}. The correspondence between these two descriptions which
is called the Cardy's condition \cite{cardy} gives a very crucial
consistency condition if one would verify the existence of some
complicated D-branes such as non-BPS D-branes discussed in \cite{sen16}
and D-branes in orbifold models or Gepner models discussed in \cite{dg}\cite{rs}\cite{douglas5}\cite{gs}. 

Another situation where the boundary state description of D-branes plays an important role is the investigation of string creation mechanism as already 
discussed in the case of flat space \cite{bergman2}. 

In this paper we discuss the string creation phenomenon in the case of
the orbifold limit of ALE spaces and show that the Witten index of ALE
space part of the open string CFT plays crucial role. So the result will 
not change if we blow up the orbifold. Further we derive the 1-loop monodromy of the quiver gauge theory on D-branes at an orbifold singularity. 

This paper is organized as follows. In section 2, we review the description of fractional D-branes on the orbifold limit of ALE spaces and some mathematical facts. In section 3, we explain already known boundary state formalism of orbifold theory and give complete proof that this is indeed the case. In section 4, we 
calculate the open string Witten index by using boundary state formalism. In section 5, we investigate the string creation phenomenon of D-branes on ALE spaces from the viewpoint of both the CFT calculation and anomaly inflow. In section 6, we consider the quiver gauge theories on D-branes and show 1-loop monodromies are equivalent to the string creation phenomenon. In section 7, we summarize the conclusions and comment on future directions. In the appendix A, we give a explicit construction of boundary states of fractional D-branes at $A_{N-1}$ singularity.

\section{Fractional D-branes at orbifold singularities}
\setcounter{equation}{0}

Here we summarize the description of the fractional D-branes at the orbifold
 $\mb{C^2/\Gamma}$ singularities \cite{douglas1}\cite{J}. We concentrate on
 the orbifold limit of ALE spaces in this and the next section. But as we 
 will see later, the main parts of the results of this paper remain correct 
 if we blow up the orbifold.
 
The orbifold singularities $\mb{C^2/\Gamma}$ are classified by the discrete
 group $\mb{\Gamma}$ falling into $A,D,E$ series. Irreducible representations 
 of $\mb{\Gamma}$ are finite dimensional (in particular one dimensional for $A$ series) and we write them as $\{\rho_i\}\ (1 \le i \le r_{\Gamma})$ ,where we define $r_{\Gamma}$ as the number of irreducible representations.
 
 Explicit $\mb{\Gamma}$-actions on $\mb{C^2}$ whose coordinates are written as  $(z_1,z_2)$ are given in \cite{douglas1}\cite{J} and for $A_{N}$ type (${\mb{\Gamma}}={\mb{Z_{N+1}}}$) they are as follows
\begin{eqnarray}
{\mb{\Gamma}}=\{1,g,\ddd,g^{N}\}\ \ \ (g^{N+1}=1) \nonumber \\
g\ :\ z_1\to e^{\f{2\pi i}{N+1}}z_1\ ,\ z_2\to e^{-\f{2\pi i}{N+1}}z_2.
\end{eqnarray}
 
This two dimensional representation is fundamental and is called as the
"natural representation" $\rho_{nat}$.
 
The geometric meaning of $\mb{C^2/\Gamma}$-singularities is that some 2-cycles
vanish at the origin and if we blow up those singularities we can define
  their intersection numbers $C_{\ap\beta}$ for 2-cycle $[\ap]$ 
and $[\beta]$.
  The following two mathematical facts are very useful.
\begin{itemize}
	\item There is a one-to-one correspondence between homology
	      2-cycles and the irreducible representations :$[\ap]\lr \rho_{\ap}$. Further in the decomposition $\rho_{\ap}\otimes \rho_{nat} =\oplus_{\beta=1}^{r_{\Gamma}}a_{\ap\beta}\rho_{\beta}$ we can identify $C_{\ap\beta}=-2\delta_{\ap\beta}+a_{\ap\beta}$ . It should also be mentioned that 2-cycles $\{[\ap]\}\ \ (1 \le \ap \le r_{\Gamma})$ are not all independent and follow the relation : $\sum_{\ap=1}^{r_{\Gamma}}n^{(0)}_\ap[\ap]\sim 0$, where we set $n^{(0)}_{\ap}=dim(\rho_{\ap})$ .
	\item $-2\delta_{\ap\beta}+a_{\ap\beta}(=C_{\ap\beta})$ are equal to the Cartan matrices of the corresponding extended Lie algebras : $\hat{\mb{\Gamma}}=\{\hat{A}_N,\hat{D}_N,\hat{E}_6,\hat{E}_7,\hat{E}_8\}$. This highly nontrivial relation between the irreducible representations of $\mb{\Gamma}$ and extended Lie algebras is called as McKay correspondence\cite{mc}.
	
\end{itemize}

Physical models at $\mb{C^2/\Gamma}$-singularities are obtained from the conventional orbifold method and further we can consider D-branes in this theory. The orbifold CFT's are not really singular because there are twisted NSNS-B fields\cite{asp} on each of 2-cycles and wrapped strings and D-branes become massive. The explicit values of B-fields are given\cite{bi}\cite{ads} as 
\begin{eqnarray}
\int_{[\ap]}B_{NS}=2\pi\f{n_{\ap}^{(0)}}{|\Gamma|}\ \ \ (n^{(0)}_{\ap}=dim(\rho_{\ap})).
\end{eqnarray}

A Dp-brane systems on $\mb{C^2/\Gamma}$ space are classified by the representations of $\mb{\Gamma}$ action on their Chan-Paton factors $\Lambda$ and we describe them as 
$\rho=\oplus_{\ap=1}^{r_{\Gamma}}n_{\ap}\rho_{\ap}$. A Dp-brane
corresponding to $\rho_{\ap}$ is called a ($\ap$-type) fractional
Dp-brane and can be interpreted as a D(p+2)-brane wrapped on a vanishing 2-cycle $[\ap]$ as discussed in \cite{douglas2}. We can always decompose a general Dp-brane system into fractional Dp-branes.
 
The spectrums of open strings attached to such a Dp-brane system are given as follows \cite{douglas1}. We define $P_{g}$-action on $\ap$-$\beta$ string, which means the open string stretched between $\ap$-type and $\beta$-type fractional Dp-brane as
\begin{eqnarray}
P_{g}\ :\ \Lambda\ \to\ \gamma^{\ap}(g)\Lambda\gamma^{\beta}(g)^{-1}\cdot R(g),
\end{eqnarray}
 where we define $\gamma^{\ap}(g)$ as the action $g\in {\mb{\Gamma}}$ on the 
Chan-Paton factor of $\rho_{\ap}$ representation and $R(g)$ as the action $g\in {\mb{\Gamma}}$ on the oscillator part of the open strings.
 
If we concentrate on the massless modes, then we get 4D N=2 super Yang-Mills theory  for p=3 and summarized as a quiver diagram\cite{douglas1}. Its gauge groups 
are $\prod_{\ap=1}^{r_{\Gamma}}U(n_{\ap})$ and hyper multiplets belong to 
$\oplus _{\ap,\beta}a_{\ap,\beta}(\bar{{\mb{n_\ap}}},{\mb{n_\beta}})$ 
representation. And the
 geometrical blowing up modes correspond to FI-terms in this Yang-Mills theory. 
 
In particular, a "bulk" D-brane corresponds to the regular representation, which is defined as $\rho_{reg}=\oplus_{\ap}n^{(0)}_{\ap}\rho_{\ap}$. The 4D N=2 super Yang-Mills theory on such branes have vanishing $\beta$-function implying superconformal invariance if we ignore each U(1) factor as discussed in \cite{ads}. 

Generally, using DBI-action and WZ-terms of D-branes we can identify\cite{ads} the complex gauge coupling of each $U(n_{\ap})$ part as
\begin{eqnarray}
\tau_{\ap}=\f{\tau}{2\pi}\int_{[\ap]}B_{NS}+\f{1}{2\pi}\int_{[\ap]}B_{R}\ ,\ 
\tau=ie^{-\phi}+\chi , \label{eqn:22}
\end{eqnarray}
 where $B_{NS}$,$B_{R}$ means NSNS,RR 2-form field and $\phi$,$\chi$ means dilaton,axion.

In the case of the regular representation, we get $\tau_{\ap}=\f{n^{(0)}_{\ap}}
{N}\tau$=const .

\section{The boundary states of D-branes at orbifold singularities}
\setcounter{equation}{0}

The boundary states of fractional D-branes at general orbifold singularities have already been constructed in \cite{dg}, but it seems that no complete proof have been given until now. So we give the proof here in the case of $\mb{C^2/\Gamma}$ orbifold\footnote{As is easily seen, generalizations of this proof to other 
orbifold cases are straightforward. }.

First we define the character $\chi^{\ap}(g)$ of g-action on the $\rho_{\ap}$ representation as $\chi^{\ap}(g)=Tr\ \gamma^{\ap}(g)$ .

The boundary states of orbifold theories generally exist for each twisted sector. Let us define $|T_g>\ \ (g\in \mb{\Gamma})$ as
\begin{eqnarray}
2\int_0^{\infty}\f{dt}{2t}Tr^{open}_{NS-R}\f{g}{|\Gamma |}\f{1+(-1)^F}{2}
e^{-2\pi tH_o}=<T_g|\Delta|T_g> ,
\end{eqnarray}
 where $Tr$ means trace over zero modes and oscillators, and $\f{1+(-1)^F}{2}$ and $H_o$ correspond to the GSO-projection and the Hamiltonian of the open string. And we defined the closed string propagator $\Delta$ as
\begin{eqnarray}
\Delta=\pi\al\int_{0}^{\infty}dl e^{-2\pi lH_c}\ \sim \f{1}{k^2}+\ddd ,
\end{eqnarray}
 where $H_c$ means the Hamiltonian of the closed string.

Note that $|T_g>$ is composed of NSNS-sector and RR-sector boundary state as
 $|T_g>=|T_g>_{NSNS}+|T_g>_{RR}$ and for each p (p=even in IIA, p=odd in IIB)
 we can construct corresponding $|T_g>$. In appendix A we give the
 explicit forms of $|T_g>$ for $A_N$-type orbifold \footnote{The similar
 calculations were carried out previously in ref \cite{billo}.} .

Every boundary state can be written as a linear combination of $\{|T_g>\}\ \ g\in \mb{\Gamma}$ . Its coefficients are determined by the Cardy's condition\cite{cardy} which requires that the $\mb{\Gamma}$-projections are well defined in open string sector.

So let us write down the Cardy's condition in the case of the open string 
stretched between fractional Dp-branes. Let us call the boundary state of a $\ap$-type fractional D-brane as $|Dp(\ap)>$.
The correct $\mb{\Gamma}$-projection for $\ap-\beta$ string is given as follows.
\begin{eqnarray}
& & 2\int_0^{\infty}\f{dt}{2t}\tilde{Tr}^{open}_{NS-R}[ \f{1+(-1)^F}{2}{\ca{P}}\ 
e^{-2\pi t H_o} ]= <Dp (\ap)|\Delta|Dp (\beta)>,  \nonumber \\
 \label{eqn:31}   \\
& & {\ca{P}}=\sum_{g\in \Gamma}\ \f{P_g}{|\Gamma|}=\sum_{g\in \Gamma}\ 
\f{P^{CP}_g\cdot g}{|\Gamma|}, \nonumber \\
& & P^{CP}_g\ :\ \Lambda\ \to\ \gamma^\ap(g)\Lambda\gamma^\beta(g)^{-1}, \nonumber 
\end{eqnarray}
 where $\tilde{Tr}$ means trace over Chan-Paton factor in addition to zero modes and oscillators. 

If we first take the trace over only Chan-Paton factor, then we get 
\begin{eqnarray}
(L.H.S)=2\int_0^{\infty}\f{dt}{2t} {Tr}^{open}_{NS-R}[(\f{1+(-1)^F}{2})
(\sum_{g\in \Gamma}\ \f{g\chi^{\ap *}(g)\chi^{\beta}(g)}{|\Gamma|})e^{-2\pi t H_o} ].
\end{eqnarray}
For example, in the case of $A_N$ series we can simplify this as follows.
\begin{eqnarray}
(L.H.S)=2\int_0^{\infty}\f{dt}{2t} {Tr}^{open}_{NS-R}[\f{1+(-1)^F}{2}
\sum_{i=0}^{N}\f{(g\xi^{\beta-\ap})^i}{N+1}e^{-2\pi t H_o} ], 
\end{eqnarray}
 where we defined $\xi=exp(\f{2\pi i}{N+1})$ and we regarded $g$ as the
 generator of $A_N={\mb{Z_{N+1}}}$.

Now, we can write down $|Dp(\ap)>$ solving the equation (\ref{eqn:31}) as
\begin{eqnarray}
|Dp(\ap)>=\sum_{g \in \Gamma}\ \chi^{\ap}(g)|T_g>.
\end{eqnarray}

This boundary state\cite{dg} represents a fractional Dp-brane which belongs to 
$\rho_{\ap}$ representation and can be interpreted as a D(p+2)-brane wrapped on vanishing $[\ap]$-cycle.

The tension and RR-charges of such a Dp-brane are easily obtained from this boundary state. To see this, we consider a bulk D-brane which belongs to the regular representation. The corresponding boundary state is 
\begin{eqnarray}
|Dp(reg)>&=&\sum_{\ap,g}\ n^{(0)}_{\ap}\chi^{\ap}(g)|T_g> \nonumber \\
         &=&\sum_{\ap}\chi^{\ap *}(1)\chi^{\ap}(g)|T_g> \nonumber \\
         &=&|\Gamma||T_1>.
\end{eqnarray}
Here we have used the following character formula 
\begin{eqnarray}
& &\sum_{\ap\in irr.rep.}\chi^{\ap *}({\ca{C}}_i)\chi^{\ap}({\ca{C}}_j)=
\f{|\Gamma|}{h_i}\delta_{ij}, \\
\end{eqnarray}
 where $h_i$ is the number of elements which belong to the
conjugacy class ${\ca{C}}_i \ \ \ (i=1\sim r_{\Gamma}) $.

Using the technique developed in \cite{dv}, we can conclude that a fractional 
Dp-brane of $\ap$-type has $\f{1}{|\Gamma|}$ tension of a bulk Dp-brane and 
its $g$-th twisted sector RR-charges are $\f{\chi^{\ap}(g)}{|\Gamma|}$.

\section{Open string Witten index}
\setcounter{equation}{0}

Here we concentrate on the internal 4D ($\mb{C^2/\Gamma}$) CFT part $|B_\ap>$ of the full boundary state $|Dp(\ap)>$ and consider the open string Witten index.
This 4D part contains all informations about the geometry of the orbifold limit
 ALE space. If we blow up the orbifold, then we will get different physical amplitudes, but topological quantities such as Witten index will not change.

Following \cite{douglas5}\cite{douglas3}, let us define the open string Witten index : $tr_{\ap\beta}\ (-1)^F$ as
\begin{eqnarray}
tr_{\ap\beta}\ (-1)^F&=&Tr^{\ap-\beta string}_{R}[(-1)^F\ e^{-2\pi tH_o}] \nonumber \\
    &=&Tr^{open}_{R}[(-1)^F\sum_{g\in \Gamma}\ \f{g\chi^{\ap *}(g)\chi^{\beta}(g)}{|\Gamma|}\ e^{-2\pi tH_o}] .
\end{eqnarray}

Using the Cardy's condition, we get 
\begin{eqnarray}
2\int_{0}^{\infty}\f{dt}{2t}tr_{\ap\beta}\ (-1)^F=<B_\ap,+|\Delta|B_\beta,->_{RR},
\end{eqnarray}
where $+,-$ means the spin structure of the boundary state (see appendix A or \cite{green1}).

As usual, massive modes (non chiral primary states) have no effect on this Witten index. So we can explicitly calculate this as follows
\begin{eqnarray}
tr_{\ap\beta}\ (-1)^F&=&-2\sum_{g\in \Gamma}\f{\chi^{\ap *}(g)\chi^{\beta}(g)}{|\Gamma|}+\sum_{g\in \Gamma}\f{\chi^{\ap *}(g)\chi^{\beta}(g)\chi^{nat}(g)}{|\Gamma|} \nonumber \\
   &=& -2\delta_{\ap\beta}+a_{\ap\beta}  \nonumber \\
   &=& C_{\ap\beta}.
\end{eqnarray}
Here we have used the decomposition
\begin{eqnarray}
\rho_{\ap}\otimes \rho_{nat} =\oplus_{\beta=1}^{r_{\Gamma}}a_{\ap\beta}\rho_{\beta}
\end{eqnarray}
 and the following formulas
\begin{eqnarray}
\sum_{g\in \Gamma}\chi^{\ap *}(g)\chi^{\beta}(g)=|\Gamma|\delta_{\ap\beta},  \\
\chi^{\rho\otimes\rho'}(g)=\chi^{\rho}(g)\cdot \chi^{\rho'}(g).
\end{eqnarray}

\section{String creation from D-branes on ALE spaces}
\setcounter{equation}{0}

\subsection{CFT calculations}

Here we investigate D2-D6 system each wrapped on the 2-cycle of ALE spaces and show the anomalous creation of string in that system. In our orbifold description such a system is regarded as fractional D0-D4 system.

First let us calculate the amplitudes of $D0-D4$ and $\bar{D}0-D4$ string. $\bar
{D}0$ means anti D0-brane and its boundary state has an extra minus sign in front of the RR-sector boundary state. We assume that a fractional $D0$ or $\bar{D}0$-brane is $\ap$-type and a fractional $D4$-brane is $\beta$-type.

Then we get 
\begin{eqnarray}
& &<D0(\ap)|\Delta|D4(\beta)> \nonumber \\
& &\ \ =-2\int_{0}^{\infty}\f{dt}{2t}\sum_{g\in \Gamma}\ \chi^{\ap *}(g)\chi
^{\beta}(g)Tr[\f{1\pm (-1)^F}{2}\f{g}{|\Gamma|}\ e^{-2\pi H_ot}] \nonumber \\
& &\ \ =-V_1\int_{0}^{\infty}\f{dt}{2t}\sum_{g\in \Gamma}\ \chi^{\ap *}(g)\chi
^{\beta}(g)\int(\f{dk}{2\pi})e^{-\f{y^2t}{2\pi\al}-2\pi\al tk^2}\hat{Tr}^{open}
_{NS-R}[(1\pm (-1)^F)\f{g}{|\Gamma|}e^{-2\hat{H_o}t}], \nonumber \\ 
\label{eqn:511}
\end{eqnarray}
 where we choose $-,+$ sign for
$D0$-brane,$\bar{D}0$-brane,respectively and superscript $\hat{*}$ means the elimination of bosonic zero modes. $V_1$ means the volume of D0-brane and $y$ is the distance between D0 and D4. 

From these amplitudes we can read off the potential energy $V$ of D0-D4 system.
Naively, one may expect that $V$ should vanish because of supersymmmetry in 
this system. To see unbroken supersymmetries, let us define by $(x^0,x^1,x^2,x^3,x^4)$ the world volume of D4-brane and by $(x^6,x^7,x^8,x^9)$ the ALE space. 
Further we write $\Gamma$-matrices and a linear combination of super-charges as $\{\Gamma_\mu\}$ and $\ep_L Q_L+\ep_R Q_R$ . Then the unbroken supersymmetries are given as
\begin{eqnarray}
D4&:& \ep_L=\g^0\g^1\g^2\g^3\g^4\ep_R \nonumber \\
D0,\bar{D}0&:& \ep_L=\pm\g^0\ep_R \nonumber \\
ALE&:& \ep_L=-\g^6\g^7\g^8\g^9\ep_L\ ,\ \ep_R=-\g^6\g^7\g^8\g^9\ep_R . \nonumber \\
\end{eqnarray}

So we get unbroken $\f{1}{8}$ supersymmetry (= 4 SUSY) .

But explicit CFT calculation of (\ref{eqn:511}) shows that such an expectation
 is partly wrong as follows
\begin{eqnarray}
& &V(\bar{D}0-D4)-V(D0-D4) \nonumber \\ 
& &\ =\f{1}{V_1}<\bar{D}0(\ap)|\Delta|D4(\beta)>-\f{1}{V_1}<D0(\ap)|\Delta|D4(\beta)>
 \nonumber \\
& &\ =\int_{0}^{\infty}\f{dt}{t}\sum_{g\in \Gamma}\chi^{\ap}(g)\chi^{\beta}(g)
\f{1}{\s{8\pi^2\al t}}e^{-\f{y^2 t}{2\pi\al}}\ \hat{Tr}_{R}(-1)^F\f{g}{|\Gamma|}
\ e^{-2\hat{H_o}t} \nonumber \\
& &\ =\int_{0}^{\infty}\f{dt}{t}\f{1}{\s{8\pi^2\al t}}\ e^{-\f{y^2 t}{2\pi\al}}
tr_{\ap\beta}(-1)^F  \nonumber \\
& &\ =-\tau_{F1}C_{\ap\beta}\ y\ \ \ (\tau_{F1}=\f{1}{2\pi\al}). \label{eqn:512}
\end{eqnarray}
In this calculation we have used the fact $\hat{Tr}_{NS}\ (-1)^F=0$ (because of 
fermionic zero modes) and $tr_{\ap\beta}(-1)^F=C_{\ap\beta}$ as shown in previous section.

Further, we argue 
\begin{eqnarray}
V(D0-D4)=0 \label{eqn:5154}.
\end{eqnarray}
(\ref{eqn:5154}) can be easily verified in the lowest order and will be correct due to the supersymmetry.

So we get 
\begin{eqnarray}
V(\bar{D}0-D4)=-\tau_{F1}C_{\ap\beta}\ y,
\end{eqnarray}
and this implies that $\bar{D}0-D4$ system is not supersymmetric. This
situation is very similar to that of $\bar{D}0-D8$ system discussed in
\cite{bergman2} and the string creation phenomenon is expected. Indeed
if we interpret $\bar{D}0-D4$ system as the rotated D0-D4 system by $\pi$ in the plane spanned by the direction of D0-brane and the transverse direction $x^5$ , then the energy difference (\ref{eqn:512}) shows that $C_{\ap\beta}$ fundamental strings are created when a D0-brane of $\ap$-type crosses a D4-brane of $\beta$-type\footnote{If $C_{\ap\beta}  <  0$, then it means that the string disappears. }.This result will be the same if we T-dualize D0-D4 system into Dp-D(4-p) system.
 
It is important to note that in this calculation the open string Witten index 
is crucial and if we blow up the orbifold into various ALE spaces or more 
globally K3 surfaces, the result will not change. So we get the following 
conclusion taking the T-duality equivalence into consideration. 

When a Dp-brane wrapped on a 2-cycle ${[\ap]}$ crosses a D(8-p)-brane 
wrapped on a 2-cycle ${[\beta]}$, $C_{\ap\beta}$ fundamental strings are created, where $C_{\ap\beta}$ is the intersection number of ${[\ap]}$ and ${[\beta]}$.

\subsection{Anomaly inflow argument}

Generally, string creation can also be investigated by using anomaly inflow argument developed in \cite{douglas4}. Anomalous creation of strings occurs if there are chiral fermions in the D-brane system such as $D5-D5$ ($D5\cap D5=$1-brane) in flat space, which is equivalent by T-duality to $D0-D8$ system.

Let us apply this method to our case : D2-brane wrapped on $[\ap]$ cycle
and D6-brane wrapped on $[\beta]$ cycle. First we T-dualize this D2-D6
system into D3-D5 system, then chiral fermions in T-dualized D3-D5
system are generated from the intersection of D-branes and the number of
chiral fermions is $C_{\ap\beta}$ including the sign of the
orientation. So we get the following anomaly equation.
\begin{eqnarray}
\Delta N=C_{\ap\beta}R\int dt\ (\f{dA^5_{(D3)}}{dt}-\f{dA^5_{(D5)}}{dt}) \ \ \ (t=x^0), 
\end{eqnarray}
 where we have compactified $x^5$ direction (radius=$R$) and $A_{D3},A_{D5}$ means the gauge field on the D3,D5-brane.

Next we take T-dual in the $x^5$ direction again and we get
\begin{eqnarray}
\Delta N=\f{C_{\ap\beta}}{2\pi R'}\int dt\ \f{d(\Delta X_5)}{dt}\ \ \ (t=x^0), 
\end{eqnarray}
 where the radius of $x^5$ direction is $R'=\f{\al}{R}$ and $\Delta X_5$ means the distance between D2 and D6.

This result is the same as that obtained in previous subsection by explicit CFT calculation. It will be important to note that the resolution by B-field is equivalent to the geometric resolution in the string creation phenomenon as the comparison between these two methods shows.

\section{1-loop monodromy of quiver gauge theories}
\setcounter{equation}{0}

\subsection{String creation and monodromy}

Here we consider fractional coincident D3-branes on the orbifold limit of ALE spaces in TypeIIB string theory. We call the number of each type of fractional D3-branes as $\{n_\ap\}$. The quiver gauge theory on that system becomes N=2 super Yang-Mills theory and so we consider 1-loop $\beta$-functions for each 
gauge group $U(n_\ap)$. We define the complex gauge couplings as 
\begin{eqnarray}
\tau_\ap=i\f{4\pi}{g^2_\ap}+\f{\theta_\ap}{2\pi}. \label{eqn:617}
\end{eqnarray}
Remember that these gauge couplings are identified as the TypeIIB background 
given in (\ref{eqn:22}).

We assume D3-brane world volume and 4D orbifold space extends in the direction of $(x^0,x^1,x^2,x^3)$ and $(x^6,x^7,x^8,x^9)$. The remaining $(x^4,x^5)$-plane will play the role of 'u-plane' or moduli space of the Coulomb branch.
 
Now let us apply the string creation mechanism discussed in previous section to
 D1-D3 system, where a fractional D1-brane stretches in the direction of $(x^0,x^4)$. If a D1-brane of $\ap$-type crosses the D3-brane system, then $\sum_{\beta=1}^{r_{\Gamma}}\ C_{\ap\beta}n_{\beta}$ fundamental strings are created.
 
This means the following monodromy for each $\ap$ around $(x^4,x^5)=(0,0)$, where the D3-branes are coincident
\begin{eqnarray}
\tau_\ap&\to&\tau_\ap+\sum_{\beta=1}^{r_{\Gamma}}\ C_{\ap\beta}n_{\beta} \nonumber \\
\tau_\ap &\sim&\tau^{(0)}_\ap+\f{1}{2\pi i}\sum_{\beta=1}^{r_{\Gamma}}\ C_{\ap\beta}n_{\beta}\ log(z_\ap)\ \ \ \ \ \ (z_\ap=x^4+ix^5). \label{eqn:613}
\end{eqnarray}

This result agrees with that of \cite{kn}, where the large N limit of the same system is treated by using AdS/CFT correspondence. And it also matches the 1-loop $\beta$-function from field theory calculations if we ignore each U(1) factor 
\begin{eqnarray}
\f{\de\tau_{\ap}(E)}{\de (logE)}=\f{1}{2\pi i}\sum_{\beta=1}^{r_{\Gamma}}\ C_{\ap\beta}n_{\beta},
\end{eqnarray}
 where we have used the identification of $z_{\ap}$ as the energy scale for $U(n_\ap)$ gauge theory.

The justification of ignoring each U(1) factor will be the same as that discussed in \cite{witten}, which is T-dual to our setup on $A_N$ series ALE spaces. In other words, the geometrical resolution of the orbifold breaks each U(1) gauge by the existence of FI-terms.

\subsection{1-loop $\beta$-function from D(-1)-D3 amplitude}

Another way to see these 1-loop monodromies is to calculate the interaction 
between a fractional D(-1)-brane and the D3-brane system. This amplitude 
corresponds to the leading correction to the action of a D(-1)-brane in the presence of D3-branes as follows.
\begin{eqnarray}
S_{inst}=2\pi i\tau_\ap=2\pi i\tau'^{(0)}_\ap - <D(-1)|\Delta|D3>+\ddd
\end{eqnarray}
Explicit calculation shows\footnote{D(-1)-D3 system is formally (time like) T-dual to D0-D4 system discussed before and the calculation is almost the same.}
\begin{eqnarray}
<D(-1)(\ap)|\Delta|D3(\beta)>&=&0 \nonumber \\
<\bar{D}(-1)(\ap)|\Delta|D3(\beta)>&=&\int_{0}^{\infty}\f{dt}{t}\ e^{-\f{|z_\ap|^2 t}{2\pi\al}}tr_{\ap\beta}\ (-1)^F \nonumber \\
  &\sim& const-2C_{\ap\beta}\ log|z_\ap|. \label{eqn:621}
\end{eqnarray}
But there is a subtlety here because there are two different 2D harmonic
 functions : $\sim log|z|$ and $\sim arg(z)$. The NSNS-scalar interaction belongs to the former and the RR-scalar the latter. So we should modify (\ref{eqn:621})
 as follows
\begin{eqnarray}
<D(-1)(\ap)|\Delta|D3(\beta)>_{NSNS}&=&-C_{\ap\beta}log|z_\ap|+const, \nonumber \\
<D(-1)(\ap)|\Delta|D3(\beta)>_{RR}&=&-iC_{\ap\beta}\ arg(z_\ap)+const, \nonumber \\
<\bar{D}(-1)(\ap)|\Delta|D3(\beta)>_{NSNS}&=&-C_{\ap\beta}log|z_\ap|+const, \nonumber \\
<\bar{D}(-1)(\ap)|\Delta|D3(\beta)>_{RR}&=&iC_{\ap\beta}\ arg(z_\ap)+const.
\end{eqnarray}
So we get 
\begin{eqnarray}
S_{inst}=2\pi i\tau_\ap=2\pi i\tau^{(0)}_\ap+(\sum_{\beta}C_{\ap\beta}n_{\beta})log(z_\ap),
\end{eqnarray}
matching the previous result (\ref{eqn:613}).

Note that this derivation of the background around D3-branes can be seen as the
 generalization of the method developed in \cite{dv}. To see this, take the low energy limit, and we can regard $|D(-1)(\ap)>$ as the vertex operator of twisted NSNS and RR scalars. But it is important that in our case the result is exact 
if we ignore the nonperturbative effects, indicating the well known fact that
 4D N=2 gauge theories receive no higher loop corrections.
 
\subsection{Comments on some other aspects of quiver gauge theories}

First we consider instantons in these gauge theories. A D-instanton of $\ap$-type
 can be regarded as an instanton of $U(n_\ap)$ gauge theory. Let us define the
 instanton numbers as
\begin{eqnarray}
\f{1}{32\pi^2}\int F_\ap\tilde{F}_\ap=k_{\ap}.
\end{eqnarray}
The bosonic moduli of this instanton corresponds to the bosonic zero modes in NS-sector D(-1)-D3 open string. So we get 
\begin{eqnarray}
dim{\ca{M}}_{instanton}=4\sum_{\ap=1}^{r_\Gamma}k_\ap n_\ap
\end{eqnarray}
as expected.
Unfortunately, as for the fermionic zero modes we get into a puzzle because 
the number of zero modes in $D(-1)-D3$ is different from that in $\bar{D}(-1)-D3$. So we leave this as an open problem.

Next we consider chiral transformations. We define $\ap$-type chiral transformation which is analogous to R-symmetry as
\begin{eqnarray}
& &\Phi_\ap\to e^{2i\phi}\Phi_\ap\ ,\ W_\ap \to e^{i\phi}W_\ap\ ,\ \theta\to e^{-i\phi}\theta \nonumber \\
& &H_{\ap\beta}\to H_{\ap\beta}\ ,\ \tilde{H}_{\ap\beta}\to \tilde{H}_{\ap\beta},\end{eqnarray}
 where $(W_\ap,\Phi_\ap)$ and $(H_{\ap\beta},\tilde{H}_{\ap\beta})$ are chiral super field descriptions of vector multiplet ($U(n_\ap)$ gauge) and hyper multiplet ($(\bar{\mb{n_\ap}},{\mb{n_\beta}})$ rep.) and $\theta$ denotes the superspace fermionic coordinate. Here we restrict the value of $\beta$ such that $C_{\ap\beta}\neq 0$ and so this chiral transformation defined for each $\ap$ is different from the conventional U(1) R-symmetry.

For $\phi=\pi$, this transformation becomes a classical symmetry of the quiver theory, but incorporating quantum effects this transformation changes the whole action as $S\to S+\Delta S$. Conventional anomaly argument shows
\begin{eqnarray}
& &\de_\mu J^\mu_{R\ap}=-\f{2}{32\pi^2}(\sum_{\beta}C_{\ap\beta}n_{\beta})F\tilde{F}\nonumber \\
& &\to\ \Delta S=\pi(\sum_{\beta}C_{\ap\beta}n_{\beta})\f{1}{16\pi^2}\int F\tilde{F}.
\end{eqnarray}

So we get the following monodromy for each $\ap$
\begin{eqnarray}
\Phi_\ap\to e^{2\pi i}\Phi_\ap\ ,\ \theta_\ap\ \to\ \theta_\ap+2\pi\sum_{\beta}C_{\ap\beta}n_{\beta}.
\end{eqnarray}
This result matches (\ref{eqn:613}) indicating the identification :$\Phi_\ap\sim z_\ap$.

We conclude this section by mentioning the nonperturbative correction to this 
quiver theory from our view point. Exact solutions to these theories are already given in \cite{witten} for T-dualized model of $A_N$ case and so we only comment on what will happen to the D3-brane system nonperturbatively.

So far we assumed that the D3-branes are coincident at the origin, but
the effect of instanton corrections may separate them and generate characteristic mass scales.

Let us consider the case $\sum_{\beta}C_{\ap\beta}n_{\beta} < 0$ which includes pure N=2 SU(n) Yang-Mills theory. In this case, (\ref{eqn:613}) indicates $\tau(z_\ap)\to -i\infty\ (z_\ap\to 0)$ and the interpretation (\ref{eqn:617}) does not work. This means that in this case D3-branes cannot be coincident and should 'fragment' by the instanton effects until the condition $\sum_{\beta}C_{\ap\beta}n_{\beta} \ge 0$ is satisfied . Quite recently some dynamics of these situations are considered from the viewpoint of AdS/CFT correspondence \cite{pol2}.

\section{Conclusions}
\setcounter{equation}{0}

In this paper we have investigated the boundary state descriptions of fractional D-branes at orbifold singularities and calculated some amplitudes of the open sting stretching between such D-branes which have the topological nature and are proportional to the open string Witten index.

Such amplitudes show the occurrence of anomalous string creation when a fractional Dp-brane crosses a fractional D(4-p)-brane in the orbifold($\mb{C^2/\Gamma}$) Type II theory. This topological nature will not change if we blow up the orbifold and the same phenomenon will occur in TypeII string theory compactified on 
K3 surfaces.

We have also mentioned the 1-loop monodromies in the quiver gauge theories on 
fractional D3-branes. Such monodromies are equivalent to the string creation phenomenon. But the interpretation of the fermionic zero modes of D(-1)-D3 string has been left as an open problem. 

Our treatment covers all types of D3-brane systems at an orbifold singularity, but if we take the large N limit, two important possibility can be considered. One possibility is that there are infinite many bulk D-branes which belong to the regular representation ($\sum_{\beta}C_{\ap\beta}n_{\beta}=0$) and 4D N=2 superconformal field theories are realized on those systems as discussed in \cite{ads}. Another possibility is that there are many fractional D-branes which belong to one particular type ($\sum_{\beta}C_{\ap\beta}n_{\beta} \to \infty$). This case will provide us very interesting 'exotic black holes'(or repulson) and AdS/CFT correspondence setups as discussed in \cite{pol2}. The dynamical aspects of these situations should be studied systematically in the future.

It will be interesting to generalize these results to more general 
compactifications such as Calabi-Yau spaces. D-branes on particular
Calabi-Yau spaces have already been discussed using the boundary state
description in \cite{dg} for $\mb{C^3/\Gamma}$ and
\cite{douglas5}\cite{gs} for quintic surface. The string creation
phenomenon in these examples will again closely be related to the open
string Witten index and the intersection numbers and will be able to be
discussed in the same way as our case. So such investigations will
require the calculation of amplitudes at only particular points in the
moduli spaces of such Calabi-Yau spaces. It will also be interesting to
study those string creation mechanism from the viewpoint of gauge
theories on the D-branes.

\bigskip

\noindent{\large \bf Acknowledgments}

I am very grateful to T.Eguchi for valuable advises, helpful discussions
and careful reading of the manuscript. I also thank S.Terashima, M.Naka
and M.Nozaki for helpful discussions and remarks.

\appendix
\setcounter{equation}{0}
\section{Explicit boundary states of $A_{N-1}$ type orbifolds}

Here we construct fractional D0-brane boundary states $|T_g>$ of $A_{N-1}$ type orbifolds explicitly by using the Cardy's condition. We represent an element of ${\mb{\Gamma}}={\mb{Z_{N}}}$ as $k\ \ (0\le k \le N-1)$.\\

[{\bf{CFT conventions}}]
\\

Let $(z,\bar{z})$ be world sheet coordinates and $(X^0\sim X^5,Z_1,Z_2)\ (\psi^0\sim \psi^5,\eta,\xi)$ be the fields corresponding to spacetime coordinates and their super partners. As in \cite{bergman1}, we use light cone formalism and we regard $X^0,X^1$ as the light cone direction. We set $a=2,3,4,5$ below.

Then the string fields in the k-th twisted sector can be expanded as
\begin{eqnarray}
\de X^a(z)&=&-i\s{\f{\al}{2}}\sum_{m\in \mb{Z}}\ap^a_{m}z^{-m-1}\ ,\ \db X^a(\bar{z})=-i\s{\f{\al}{2}}\sum_{m\in \mb{Z}}\tap^a_{m}\bar{z}^{-m-1} \nonumber \\
\psi^a(z) &=&\sum_{r}\psi_r^a z^{-r-\f12}\ , \ \tilde{\psi}^a(\bar{z}) =\sum_{r}\tilde{\psi}_r^a \bar{z}^{-r-\f12} \nonumber \\
\de Z_1(z)&=&-i\s{\al}\sum_{m\in \mb{Z}}\ap_{m-\fkN}z^{-m-1-\fkN}\ ,\ \db Z_1(\bar{z})=-i\s{\al}\sum_{m\in \mb{Z}}\tap_{m+\fkN}\bar{z}^{-m-1+\fkN} \nonumber \\
\de \bar{Z_1}(z)&=&-i\s{\al}\sum_{m\in \mb{Z}}\bap_{m+\fkN}z^{-m-1+\fkN}\ ,\ \db \bar{Z_1}(\bar{z})=-i\s{\al}\sum_{m\in \mb{Z}}\tbap_{m-\fkN}\bar{z}^{-m-1-\fkN} \nonumber \\
\de Z_2(z)&=&-i\s{\al}\sum_{m\in \mb{Z}}\beta_{m+\fkN}z^{-m-1+\fkN}\ ,\ \db Z_2(\bar{z})=-i\s{\al}\sum_{m\in \mb{Z}}\tbe_{m-\fkN}\bar{z}^{-m-1-\fkN} \nonumber \\
\de \bar{Z_2}(z)&=&-i\s{\al}\sum_{m\in \mb{Z}}\bbe_{m-\fkN}z^{-m-1-\fkN}\ ,\ \db \bar{Z_2}(\bar{z})=-i\s{\al}\sum_{m\in \mb{Z}}\tbbe_{m+\fkN}\bar{z}^{-m-1+\fkN} \nonumber \\ 
\eta(z)&=&\s{2}\sum_{r}\eta_{r-\fkN}z^{-r-\f12+\fkN}\ , \ \eta(\bar{z})=\s{2}\sum_{r}\tet_{r+\fkN}z^{-r-\f12-\fkN} \nonumber \\
\xi(z)&=&\s{2}\sum_{r}\eta_{r+\fkN}z^{-r-\f12-\fkN}\ , \ \xi(\bar{z})=\s{2}\sum_{r}\txi_{r-\fkN}z^{-r-\f12+\fkN}, \nonumber \\
\end{eqnarray}
 where the modings : $r$ of fermions are all half integer in NS-sector and integer in R-sector.
And non trivial (anti)commutation relations are as follows (we only show left-moving parts only)
\begin{eqnarray}
[\bap_{m+\fkN}\ ,\ \ap_{n-\fkN}]=(m+\fkN)\delta_{m,-n}\ &,& \ [\bbe_{m-\fkN}\ ,\ \be_{n+\fkN}]=(m-\fkN)\delta_{m,-n} \nonumber \\
\{\eta_{r-\fkN}\ ,\ \bet_{r+\fkN}\}=\delta_{r+s}\ &,& \ \{\xi_{r+\fkN}\ ,\ \bxi_{r-\fkN}\}=\delta_{r+s}. \nonumber \\
\end{eqnarray}

So k-th twisted sector vacuum is defined as 
\begin{eqnarray}
\ap_{m-\fkN}|0>_k=0\ (m > 0)\ ,\ \tap_{m+\fkN}|0>_k=0\ (m \ge 0), \nonumber \\
\bap_{m+\fkN}|0>_k=0\ (m \ge 0)\ ,\ \tap_{m-\fkN}|0>_k=0\ (m > 0), \nonumber \\
\be_{m+\fkN}|0>_k=0\ (m \ge 0)\ ,\ \tbap_{m-\fkN}|0>_k=0\ (m > 0), \nonumber \\
\bbe_{m-\fkN}|0>_k=0\ (m > 0)\ ,\ \tap_{m+\fkN}|0>_k=0\ (m \ge 0), \nonumber \\
\eta_{r-\fkN}|0>_k=0\ (r > 0)\ ,\ \tet_{r+\fkN}|0>_k=0\ (r \ge 0), \nonumber \\
\bet_{r+\fkN}|0>_k=0\ (r \ge 0 ,\ \tbet_{r-\fkN}|0>_k=0\ (r > 0), \nonumber \\
\xi_{r+\fkN}|0>_k=0\ (r \ge 0)\ , \ \txi_{r-\fkN}|0>_k=0\ (r > 0), \nonumber \\
\bxi_{r-\fkN}|0>_k=0\ (r > 0)\ , \ \tbxi_{r+\fkN}|0>_k=0\ (r \ge 0).\nonumber \\
\end{eqnarray}

Then the zero point energy is 
\begin{eqnarray}
a_{NS}=-\f12+\fkN\ , \ a_{R}=0.    
\end{eqnarray}
and physical states satisfy $(L_0+a_{sector})|phys>=(\bar{L}_0+a_{sector})|phys>=0$. Here we defined
\begin{eqnarray}
L_0&=&\f{\al p^2}{4}+\sum_{a=2}^{5}(\sum_{n\ge 1}\ap^a_{-n}\ap^a_{n}+\sum_{r > 0}r\psi^a_{-r}\psi^a_{r})\nonumber \\
& &+\sum_{n=-\infty}^{\infty}(:\ap_{n-\fkN}\bap_{-n+\fkN}:+:\be_{n+\fkN}\bbe_{-n-\fkN}:)\nonumber \\
& &+\sum_{r=-\infty}^{\infty}\{:(r-\fkN)\eta_{r-\fkN}\bet_{n+\fkN}:+:(r+\fkN)\xi_{r+\fkN}\bxi_{r-\fkN}:\},
\end{eqnarray}
and $\tilde{L_0}$ similarly. For later use, we define closed string
Hamiltonian $H_{c}$ as
\begin{eqnarray}
H_{c}=L_0+\tilde{L_0}+a^L_{sector}+a^R_{sector} .
\end{eqnarray}
and open string Hamiltonian:$H_{o}$ similarly.\\

[{\bf{Construction of boundary state}}]\\

The D0-brane k-th twisted sector boundary state:$|T_k,\ep>\ \ (0\le k \le N-1)$ in the theory of TypeIIA on $R^{1,5}\times C^2/\mb{Z_N}$  is defined as follows. 
\begin{eqnarray}
& &(\ap^a_n-\tap^a_{-n})|T_k,\ep>=0\ ,\ (\psi^a_r-i\ep\psi^a_{-r})|T_k,\ep>=0 \nonumber \nonumber \\
& &(\ap_{m-\fkN}-\tap_{-m+\fkN})|T_k,\ep>=0\ ,\ (\bap_{m+\fkN}-\tbap_{-m-\fkN})|T_k,\ep>=0 \nonumber \\
& &(\be_{m+\fkN}-\tbe_{-m-\fkN})|T_k,\ep>=0\ ,\ (\bbe_{m-\fkN}-\tbbe_{-m+\fkN})|T_k,\ep>=0 \nonumber \\
& &(\eta_{r-\fkN}-i\ep\tet_{-r+\fkN})|T_k,\ep>=0\ ,\ (\bet_{r+\fkN}-i\ep\tbet_{-r-\fkN})|T_k,\ep>=0 \nonumber \\
& &(\xi_{r+\fkN}-i\ep\txi_{-r-\fkN})|T_k,\ep>=0\ ,\ (\bxi_{r-\fkN}-i\ep\tbxi_{-r+\fkN})|T_k,\ep>=0 \nonumber \\
\end{eqnarray}
, where $\vec{p}$ means the momentum of a D0-brane in the direction of $x^1\sim x^5$.
The solution to these constraints is 
\begin{eqnarray}
|T_k,\ep,\vec{p}>_{NSNS}&=&exp[\sum_{n=1}^{\infty}(\f{1}{n}\sum_{a=2}^{5}\ap^a_{-n}\tap^a_{-n})+i\ep\sum_{r > 0}(\sum_{a=2}^{5}\p^a_{-r}\bp^a_{-r})] \nonumber  \\
  & &\times exp[\sum_{n=0}^{\infty}(\f{1}{n+\fkN}\ap_{-n-\fkN}\tbap_{-n-\fkN})+\sum_{n=1}^{\infty}(\f{1}{n-\fkN}\tap_{-n+\fkN}\bap_{-n+\fkN}) \nonumber  \\
  & &+\sum_{n=0}^{\infty}(\f{1}{n-\fkN}\be_{-n+\fkN}\tbbe_{-n+\fkN})+\sum_{n=1}^{\infty}(\f{1}{n+\fkN}\tbe_{-n-\fkN}\bbe_{-n-\fkN})  \nonumber  \\
  & &+i\ep(\sum_{r > 0}\et_{-r-\fkN}\tbet_{-r-\fkN}+\sum_{r > 0}\tet_{-r+\fkN}\bet_{-r+\fkN})            \nonumber  \\
  & &+i\ep(\sum_{r > 0}\xi_{-r+\fkN}\tbxi_{-r+\fkN}+\sum_{r > 0}\txi_{-r-\fkN}\bxi_{-r-\fkN})]|T_i,\ep,\vec{p}>^{(0)}_{NSNS}, \label{eqn:323} \nonumber \\  \\
 |T_k,\ep,\vec{p}>_{RR}&=&exp[\sum_{n=1}^{\infty}(\f{1}{n}\sum_{a=2}^{5}\ap^a_{-n}\tap^a_{-n})+i\ep\sum_{r=1}^{\infty}(\sum_{a=2}^{5}\p^a_{-r}\bp^a_{-r})] \nonumber  \\
  & &\times exp[\sum_{n=0}^{\infty}(\f{1}{n+\fkN}\ap_{-n-\fkN}\tbap_{-n-\fkN})+\sum_{n=1}^{\infty}(\f{1}{n-\fkN}\tap_{-n+\fkN}\bap_{-n+\fkN}) \nonumber  \\
  & &+\sum_{n=0}^{\infty}(\f{1}{n-\fkN}\be_{-n+\fkN}\tbbe_{-n+\fkN})+\sum_{n=1}^{\infty}(\f{1}{n+\fkN}\tbe_{-n-\fkN}\bbe_{-n-\fkN})  \nonumber  \\
  & &+i\ep(\sum_{r=0}^{\infty}\et_{-r-\fkN}\tbet_{-r-\fkN}+\sum_{r=1}^{\infty}\tet_{-r+\fkN}\bet_{-r+\fkN})            \nonumber  \\
  & &+i\ep(\sum_{r=1}^{\infty}\xi_{-r+\fkN}\tbxi_{-r+\fkN}+\sum_{r=0}^{\infty}\txi_{-r-\fkN}\bxi_{-r-\fkN})]|T_i,\ep,\vec{p}>^{(0)}_{RR}.  \label{eqn:324} 
\end{eqnarray}
Here $| T_i >^{(0)}$ means zero mode part of the boundary state. 

The GSO-projection invariant combination is
\begin{eqnarray}
|T_k>_{NSNS}&=&\f{1}{2}\int(\f{dp}{2\pi})^{5}[|T_i,+,\vec{p}>_{NSNS}\pm|T_i,-,\vec{p}>_{NSNS}], \label{eqn:321} \\
|T_k>_{RR}&=&\f{1}{2}\int(\f{dp}{2\pi})^{5}[|T_i,+,\vec{p}>_{RR}\pm|T_i,-,\vec{p}>_{RR}].  \label{eqn:322}   
\end{eqnarray}
Signs in these equations are determined\footnote{It is easily shown by requiring the GSO invariance that each of these sign is $+$ if there are fermionic zero mode or is $-$ if not. } as follows.

k=0 : (-) for NS-sector , (+) for R-sector

k=$\f{N}{2}$ : (+) for NS-sector , (+) for R-sector

Others : (-) for NS-sector , (+) for R-sector \\

Finally, $|T_k>$ is constructed from a linear combination of $|T_k>_{NSNS}$ and $|T_k>_{RR}$ as
\begin{eqnarray}
|T_k>&=&\f{{\ca{N}}^{NSNS}_k}{2}|T_k>_{NSNS}+\f{{\ca{N}}^{RR}_k}{2}|T_k>_{RR}, \label{eqn:320} 
\end{eqnarray}
 where the coefficients : ${\ca{N}}^{NSNS}_k\ ,\ {\ca{N}}^{RR}_k$ are 
determined from the Cardy's condition :(\ref{eqn:31}) as we will see below.\\

[{\bf{Boundary state calculations}}]
\\

In order to determine the coefficients in eq.(\ref{eqn:320}) we require
Cardy's condition eq.(\ref{eqn:31}) .

Before we do that, let us define "Z" functions following ref.\cite{pol1} 
as
\begin{eqnarray}
Z^\ap_\be(\tau)&=&q^{\f{3\ap^2-1}{24}}e^{i\pi\ap\be}\prod_{m=1}^{\infty}(1+e^{i\pi\be}q^{m-\f{1-\ap}{2}})(1+e^{-i\pi\be}q^{m-\f{1+\ap}{2}}),  \\
\eta(\tau)&=&q^{-\f{1}{24}}\prod_{m=1}^{\infty}(1-q^m),
\end{eqnarray}
where we set $q=e^{i2\pi\tau}=e^{-4\pi l}$
 
These functions have following modular properties
\begin{eqnarray}
Z^\ap_\be(\tau)=Z^\beta_{-\ap}(-\f{1}{\tau})\ , \ \eta(-\f{1}{\tau})=\s{-i\tau}\eta(\tau).
\end{eqnarray}

First we show the calculations of (R.H.S) of eq.(\ref{eqn:31}). If k is not 0, then we get

\begin{eqnarray}
<+,T_k|\Delta|+,T_k>_{NSNS}&=&V_1\int_{0}^{\infty}dl\ (\f{dk}{2\pi})^5\ e^{-\pi l\al k^2}\f{|Z^0_0(2il)|^2|Z^{2\fkN}_{0}(2il)|^2}{\eta(2il)^4|Z^{1-2\fkN}_1(2il)|^2},  \nonumber \\
<+,T_k|\Delta|-,T_k>_{NSNS}&=&V_1\int_{0}^{\infty}dl\ (\f{dk}{2\pi})^5\ e^{-\pi l\al k^2}\f{|Z^0_1(2il)|^2|Z^{-2\fkN}_{1}(2il)|^2}{\eta(2il)^4|Z^{1-2\fkN}_1(2il)|^2}, \nonumber  \\
<+,T_k|\Delta|+,T_k>_{RR}&=&V_1\int_{0}^{\infty}dl\ (\f{dk}{2\pi})^5\ e^{-\pi l\al k^2}\f{|Z^1_0(2il)|^2|Z^{1-2\fkN}_{0}(2il)|^2}{\eta(2il)^4|Z^{1-2\fkN}_1(2il)|^2},  \nonumber \\
<+,T_k|\Delta|-,T_k>_{RR}&=&0 \ \  \mbox{(This is because of fermionic zero modes.)}, \nonumber \\
\end{eqnarray}
where we defined $V_1$ as the volume of the Neumann direction.

Doing the Gauss integrations and modular transformations, we get
\begin{eqnarray}
& &<T_k|\Delta|T_k> \nonumber \\
& &=\f{\pi\al V_1}{8\cdot (2\pi)^5}\int_0^{\infty}dl\ (l\al)^{-\f52}\f{1}{\eta(2il)^4|Z^{1-2\fkN}_1(2il)|^2} \nonumber \\
& &\cdot\{|{\ca{N}}^{NSNS}_k|^2|Z^0_0(2il)|^2|Z^{2\fkN}_0(2il)|^2-|{\ca{N}}^{NSNS}_k|^2|Z^0_1(2il)|^2|Z^{-2\fkN}_1(2il)|^2 \nonumber \\
& &-|{\ca{N}}^{RR}_k|^2|Z^1_0(2il)|^2|Z^{1-2\fkN}_0(2il)|^2\}  \nonumber \\
& &=\f{\pi(\al)^{-\f32}V_1}{2^{\f32}(2\pi)^5}\int_0^{\infty}dt\ t^{-\f32}\f{1}{\eta(it)^4|Z^1_{2\fkN-1}(it)|^2}\cdot\{|{\ca{N}}^{NSNS}_k|^2|Z^0_0(it)|^2|Z^0_{2\fkN}(it)|^2 \nonumber \\
& &-|{\ca{N}}^{NSNS}_k|^2|Z^1_0(it)|^2|Z^1_{2\fkN}(it)|^2-|{\ca{N}}^{RR}_k|^2|Z^0_1(it)|^2|Z^0_{2\fkN-1}(it)|^2\},  \nonumber \\
\end{eqnarray}
where we defined $t=\f{1}{2l}$.

In the same way if k=0 (untwisted sector) then we get
\begin{eqnarray}
<T_0|\Delta|T_0>&=&\f{\pi\al V_1}{8}\int_0^{\infty}dl(\f{dk}{2\pi})^9\ e^{-\pi l\al k^2}\f{1}{\eta(2il)^8}\{|{\ca{N}}^{NSNS}_k|^2|Z^0_0(2il)|^4 \nonumber \\
& &-|{\ca{N}}^{NSNS}_k|^2|Z^0_1(2il)|^4-|{\ca{N}}^{RR}_k|^2|Z^1_0(2il)|^4\} \nonumber \\
&=& 2^{-\f{17}{2}}\pi^{-8}(\al)^{-\f72}V_1\int_0^{\infty}dt\ t^{-\f32}\f{1}{\eta(it)^8}\{|{\ca{N}}^{NSNS}_k|^2|Z^0_0(it)|^4  \nonumber \\
& &-|{\ca{N}}^{NSNS}_k|^2|Z^1_0(it)|^4-|{\ca{N}}^{RR}_k|^2|Z^0_1(it)|^4\}. \nonumber \\
\end{eqnarray}
\\

[{\bf{Open string calculations}}]
\\

Next we show the calculations of (L.H.S) of eq.(\ref{eqn:31}). If k is not 0, then we get
\begin{eqnarray}
& &2\int_{0}^{\infty}\f{dt}{2t}Tr_{NS-R}[\f{1+(-1)^F}{2}\ g^k\ e^{-2\pi tH_{o}}] \nonumber \\
&=&2^{-\f52}(\pi)^{-1}(\al)^{-\f12}V_1\int_0^{\infty}dt\ t^{-\f32}\f{C}{\eta(it)^4|Z^1_{2\fkN-1}(it)|^2}\{|Z^0_0(it)|^2|Z^0_{2\fkN}(it)|^2 \nonumber \\
&-&|Z^0_1(it)|^2|Z^0_{2\fkN-1}(it)|-|Z^1_0(it)|^2|Z^1_{2\fkN}(it)|\},
\end{eqnarray}
where C is 4 for $k=\f{N}{2}$ and is 1 for others.

If k is 0, then we get
\begin{eqnarray}
& &2\int_{0}^{\infty}\f{dt}{2t}Tr_{NS-R}[\f{1+(-1)^F}{2}\ e^{-2\pi tH_{o}}] \nonumber \\
&=&2^{-\f52}(\pi)^{-1}(\al)^{-\f12}V_1\int_0^{\infty}dt\ t^{-\f32}\cdot\f{Z^0_0(it)^4-Z^1_0(it)^4-Z^0_1(it)^4}{\eta(it)^8}.  \nonumber \\
\end{eqnarray}

[{\bf{Determination of the normalization}}]
\\

Now we can determine the coefficients in eq.(\ref{eqn:320}) by using Cardy's condition eq.(\ref{eqn:31}).

Then the results are as follows \\

(1)$k=0$ case : ${\ca{N}}^{NSNS}_0={\ca{N}}^{RR}_0=2^3\pi^{\f72}\al^{\f32}$, \\

(2)$k=\f{N}{2}$ case : ${\ca{N}}^{NSNS}_0={\ca{N}}^{RR}_0=2^4\pi^{\f32}\al^{\f12}$,\\

(3)$k\ne0,\f{N}{2}$ case : $2{\ca{N}}^{NSNS}_0={\ca{N}}^{RR}_0=2^3\pi^{\f32}\al^{\f12}$. \\

\end{document}